\documentclass[prb,amsmath,amsfonts,amssymb,twocolumn,floatfix,reprint]{revtex4-1}
\usepackage{graphicx}
\usepackage{bm}
\usepackage{color}

\begin{document}
\title{Projected Hartree Fock Theory as a Polynomial Similarity Transformation Theory of Single Excitations}
\author{Yiheng Qiu}
\affiliation{Department of Chemistry, Rice University, Houston, TX 77005-1892}

\author{Thomas M. Henderson}
\affiliation{Department of Chemistry, Rice University, Houston, TX 77005-1892}
\affiliation{Department of Physics and Astronomy, Rice University, Houston, TX 77005-1892}

\author{Gustavo E. Scuseria}
\affiliation{Department of Chemistry, Rice University, Houston, TX 77005-1892}
\affiliation{Department of Physics and Astronomy, Rice University, Houston, TX 77005-1892}
\date{\today}

\begin{abstract}
Spin-projected Hartree-Fock is introduced as a particle-hole excitation ansatz over a symmetry-adapted reference determinant.  Remarkably, this expansion has an analytic expression that we were able to decipher.  While the form of the polynomial expansion is universal, the excitation amplitudes need to be optimized.  This is equivalent to the optimization of orbitals in the conventional projected Hartree-Fock framework of non-orthogonal determinants. Using the inverse of the particle-hole expansion, we similarity transform the Hamiltonian in a coupled-cluster style theory. The left eigenvector of the non-hermitian Hamiltonian is constructed in a similar particle-hole expansion fashion, and we show that to numerically reproduce variational projected Hartree-Fock results, one needs as many pair excitations in the bra as the number of strongly correlated entangled pairs in the system.  This single-excitation polynomial similarity transformation theory is an alternative to our recently presented double excitation theory, but supports projected Hartree-Fock and coupled cluster simultaneously rather than interpolating between them.
\end{abstract}
\maketitle

The main difficulty in computational quantum chemistry is the need for an accurate description of electronic correlation effects.  When electrons are weakly correlated and a mean-field picture is qualitatively accurate, it is probably fair to say that this difficulty is overcome by using some form of coupled cluster (CC) theory.\cite{CoesterKummel,Paldus1999,Bartlett2007,ShavittBartlett}  When the mean-field picture is qualitatively incorrect, however, and electrons are strongly correlated, the situation is significantly different, and it is probably equally fair to say that there is no completely general solution to the problem.\cite{SimonsHubbard}  When the number of strongly correlated electrons is not too large, active-space methods work well, but ultimately these become too computationally cumbersome to be of practical utility.

One appealing approach to the strong correlation problem is the use of projected Hartree-Fock (PHF).\cite{Lowdin55c,Ring80,Blaizot85,Schmid2004,PQT,PHF}  When a system becomes strongly correlated, the mean-field tends to signal its own demise by spontaneously breaking a symmetry of the Hamiltonian.  The component of the broken-symmetry mean-field wave function which has the correct symmetries will be a multi-determinantal wave function which typically offers a fairly reliable description of the strong correlations.  Moreover, by optimizing the mean-field in the presence of the symmetry projection operator, one can \textit{deliberately} break symmetries even when those symmetries do not break spontaneously; this variation after projection approach to symmetry breaking and restoration leads to wave functions which are well-behaved as a function of Hamiltonian parameters.\cite{PQT,PHF}

\begin{figure}
\includegraphics[width=\columnwidth]{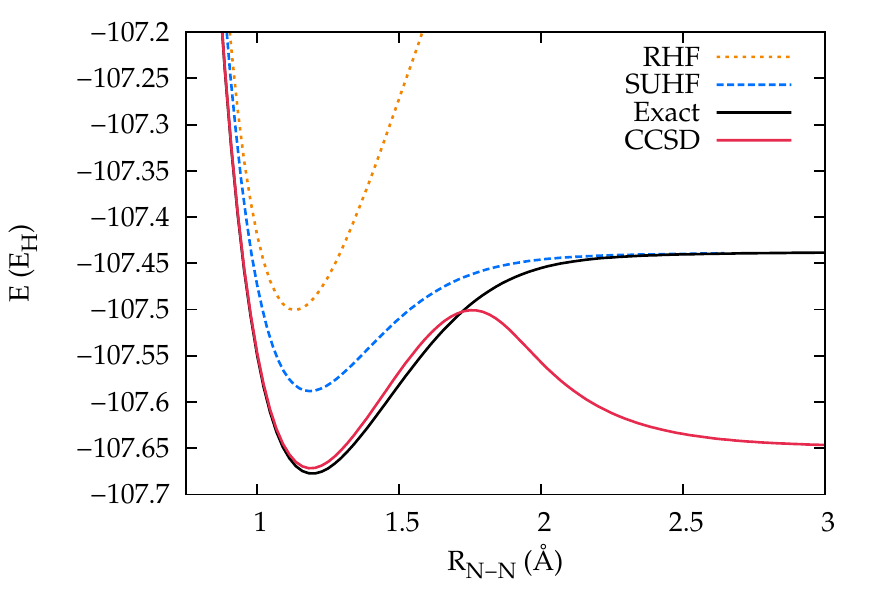}
\caption{Dissociation of N$_2$ in the STO-3G basis.
\label{Fig:N2STO3G}}
\end{figure}

To see all this in action, we show in Fig. \ref{Fig:N2STO3G} a plot of the dissociation of the N$_2$ molecule in a minimal basis, where exact results are available.  The key features we wish to emphasize are as follows.  First, the symmetry-adapted mean-field (RHF) is not terrible near equilibrium, but is useless toward dissociation.  Second, coupled cluster with single and double excitations (CCSD) based on the symmetry-adapted reference is highly accurate near equilibrium where RHF is reasonable, but as the bond stretches, RHF breaks down, and the molecule becomes more strongly correlated, CCSD goes haywire.  Finally, spin-projected unrestricted Hartree-Fock (SUHF) is well-behaved everywhere but is clearly missing a chunk of the correlation energy -- that is, it lacks an accurate accounting for the weak correlations that CCSD so readily recovers.  We note in passing that the exactness of SUHF at dissociation is a consequence of the minimal basis set and is not a general result.

\begin{figure}
\includegraphics[width=\columnwidth]{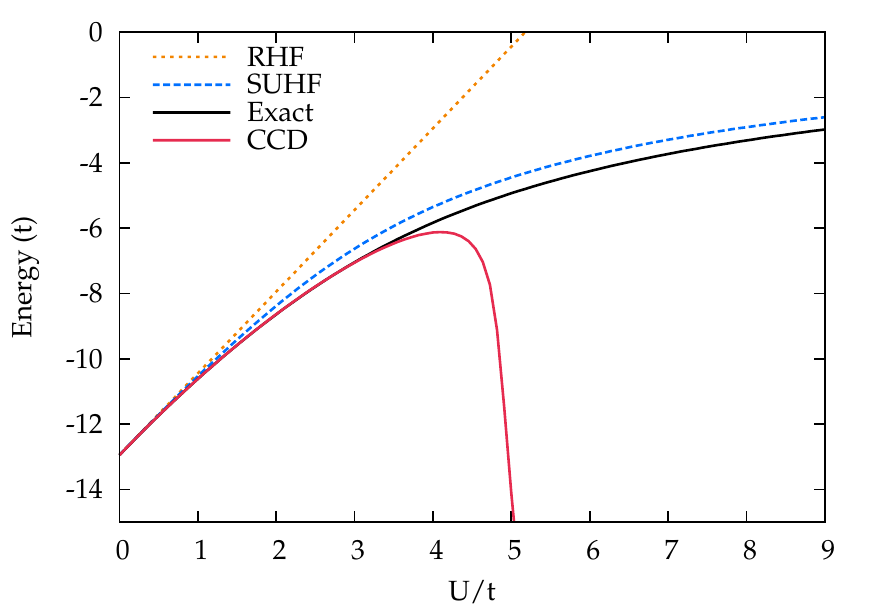}
\caption{Total energies in the 10-site half-filled periodic one-dimensional Hubbard Hamiltonian.
\label{Fig:Hubbard10}}
\end{figure}

The breakdown may be even clearer in the Hubbard Hamiltonian,\cite{Hubbard1963} as shown in Fig. \ref{Fig:Hubbard10}.  Here, we see that for small on-site repulsion $U$ where the system is weakly correlated, CCSD is exceptionally accurate, but that it breaks down completely for larger $U$; SUHF, meanwhile, is reasonable but imperfect everywhere.\cite{PHFHubbard1,PHFHubbard2}

In an ideal world, one could simply combine PHF and coupled cluster, but this task is complicated by the very different natures of the two theories.  The PHF wave function is concisely written as a (short) expansion in a set of non-orthogonal broken-symmetry determinants, while the coupled cluster wave function is written as a (long) expansion in a set of orthogonal symmetry-adapted determinants.  The PHF energy is an expectation value, and the PHF wave function is obtained using the variational principle, where coupled cluster uses a projective Schr\"odinger equation approach to define a non-variational energy.  It is difficult to see how to cleanly marry these two approaches.

In this communication, building on our work on polynomial similarity transformations (PoST),\cite{Scuseria2016} we show how one can cast PHF in terms of particle-hole excitations out of a symmetry-adapted determinant and, moreover, we show how one can optimize the PHF wave function in coupled-cluster-like fashion as opposed to the more traditional variational optimization.   We will here consider only what we call SUHF, and will further limit ourselves to projection onto singlet states ($s = 0)$ only, but the theory can be extended to a much more general framework including other quantum numbers and symmetries.

We start from the observation that an $m_S = 0$ unrestricted Hartree-Fock (UHF) determinant can be written as a Thouless transformation of an RHF determinant:
\begin{equation}
|\mathrm{UHF}\rangle = \mathrm{e}^{T_1 + U_1} \, |\mathrm{RHF}\rangle,
\label{Eqn:Thouless}
\end{equation}
where we have introduced
\begin{subequations}
\begin{align}
T_1 &= \sum t_i^a \, E_a^i,
\\
U_1 &= \sum u_i^a \, S_a^i,
\\
E_a^i &= c_{a_\uparrow}^\dagger \, c_{i_\uparrow} + c_{a_\downarrow}^\dagger \, c_{i_\downarrow},
\\
S_a^i &= c_{a_\uparrow}^\dagger \, c_{i_\uparrow} - c_{a_\downarrow}^\dagger \, c_{i_\downarrow}.
\end{align}
\end{subequations}
We follow the convention that spatial orbitals indexed $i$, $j$, $k$, \ldots ($a$, $b$, $c$, \ldots) are occupied (empty) in $|\mathrm{RHF}\rangle$. Note that the Thouless transformation above does not preserve normalization and that we have assumed that $|\mathrm{UHF}\rangle$ and $|\mathrm{RHF}\rangle$ are not orthogonal.

A key result in this paper is that the SUHF wave function, traditionally obtained via a singlet symmetry projection operator $P$ acting on $|\mathrm{UHF}\rangle$, can be written as
\begin{equation}
|\mathrm{SUHF}\rangle 
 = P \, |\mathrm{UHF}\rangle
 = \mathrm{e}^{T_1} \, F(K_2) |\mathrm{RHF}\rangle,
\label{Eqn:SUHFwfn}
\end{equation}
where $K_2$ is the singlet component of $U_1^2$,
\begin{equation}
K_2 = -\frac{1}{6} \, \sum \left(u_i^a \, u_j^b + 2 \, u_i^b \, u_j^a\right) \, E_a^i \, E_b^j,
\end{equation}
and the polynomial $F(K_2)$ is given by
\begin{subequations}
\label{Eqn:Sinh}
\begin{align}
F(K_2) &= \frac{\mathrm{sinh}(\sqrt{6 \, K_2})}{\sqrt{6 \, K_2}}
\\
 &= 1 + K_2 + \frac{3}{10} \, K_2^2 + \frac{3}{70} \, K_2^3  + \ldots.
\end{align}
\end{subequations}
Only even powers of $U_1$ appear, as odd powers of $U_1$ break spin symmetry, a fact already noted in Ref. \onlinecite{Piecuch1996}, although these authors did not point out that the PHF wave function could be expressed as a polynomial of the double-excitation operator $K_2$, a key result needed for solving for the $u_i^a$ amplitudes as we do here.

Details of the proof will be presented elsewhere, but a rough sketch of the idea proceeds as follows.  The expression for each term $P \, U_1^n$ can be obtained by analytically integrating over spin rotation angles, in a manner basically analagous to the numerical integration done in SUHF among non-orthogonal determinants, but working here with orthogonal particle-hole excitations.  That $F(K_2)$ is given by the sinh polynomial is recognized by direct inspection of the individual projected terms and numerically proven below by comparison with our previous implementation of SUHF.\cite{PHF}  Optimization of the coefficients $u_i^a$ plays the role of orbital optimization in SUHF.\cite{PHF}

Thus far, all we have done is to reparameterize the SUHF wave function.  This is an important step, but it should be noted that conventionally SUHF defines the energy as an expectation value which it variationally minimizes with respect to the UHF determinant.  In our language, this would require us to solve
\begin{subequations}
\label{Eqn:FDagHF}
\begin{align}
E &= \frac{\langle \mathrm{RHF} | F(K_2^\dagger) \, \mathrm{e}^{T_1^\dagger} \, H \, \mathrm{e}^{T_1} \, F(K_2) |\mathrm{RHF}\rangle}{\langle \mathrm{RHF} | F(K_2^\dagger) \, \mathrm{e}^{T_1^\dagger} \, \mathrm{e}^{T_1} \, F(K_2) | \mathrm{RHF}\rangle},
\\
0 &= \frac{\partial E}{\partial t_i^a} = \frac{\partial E}{\partial u_i^a},
\end{align}
\end{subequations}
which is not readily compatible with the typical approach used in traditional CC theory.

In standard CC theory, we construct a similarity-transformed Hamiltonian
\begin{equation}
\bar{H} = \mathrm{e}^{-T} \, H \, \mathrm{e}^T,
\end{equation}
where $T$ creates excitations out of the reference, which we will denote simply as
\begin{equation}
T = \sum t_\mu \, Q_\mu^\dagger
\end{equation}
where $t_\mu$ are excitation amplitudes and excited determinants $|\Phi_\mu\rangle$ are created by the action of $Q_\mu^\dagger$ on the reference.  We wish to choose the similarity transformation such that $|\mathrm{RHF}\rangle$ is a right-hand eigenstate of $\bar{H}$.  Because $\bar{H}$ is non-Hermitian, we must also solve for a left-hand eigenstate, which is parameterized as $\langle\mathrm{RHF}|(1+Z)$, where
\begin{equation}
Z = \sum z_\mu \, Q_\mu.
\end{equation}
We define the energy by the expectation value of $\bar{H}$, which is made stationary with respect to the amplitudes defining $T$ and $Z$:
\begin{subequations}
\begin{align}
E &= \langle \mathrm{RHF} | \left(1 + Z\right) \, \bar{H} | \mathrm{RHF}\rangle,
\\
0 &= \frac{\partial E}{\partial t_\mu} = \frac{\partial E}{\partial z_\mu}.
\end{align}
\end{subequations}

Our goal is thus to write PHF in a coupled-cluster-like language, because if we can do so, then combining CC and PHF is essentially straightforward.  To write PHF in this manner, we proceed in analogy with CC above and define a similarity-transformation for PHF:
\begin{equation}
\bar{H}_\mathrm{PHF} = F^{-1}(K_2) \, \mathrm{e}^{-T_1} \, H \, \mathrm{e}^{T_1} \, F(K_2).
\end{equation}
Because $T_1$ and $K_2$ are both excitation operators, they commute, so we need not worry about the order of $F(K_2)$ and $\exp(T_1)$.  Whereas the singlet projection operator $P$ does not have an inverse, the inverse polynomial $F^{-1}(K_2)$ can be extracted from a Taylor series expansion of $1/F(x)$ and begins
\begin{equation}
F^{-1}(K_2) = 1 - K_2 + \frac{7}{10} \, K_2^2 - \frac{31}{70} \, K_2^3 + \ldots.
\end{equation}
As with CC theory, we want the right-hand eigenstate of $\bar{H}_\mathrm{PHF}$ to be the symmetry-adapted determinant, but we will need a more complicated left-hand eigenstate to approximate the variational PHF case.  Thus, we define
\begin{subequations}
\label{Eqn:PSUHF}
\begin{align}
E &= \langle \mathrm{RHF} | \left(1 + Z_1\right) \, G(L_2) \, \bar{H}_\mathrm{PHF} | \mathrm{RHF} \rangle,
\\
Z_1 &= \sum z^i_a \, E^a_i,
\\
L_2 &= -\frac{1}{6} \, \sum \left(v^i_a \, v^j_b + 2 \, v^i_b \, v^j_a\right) E^a_i \, E^b_j,
\\
E^a_i &= \left(E_i^a\right)^\dagger = c_{i_\uparrow}^\dagger \, c_{a_\uparrow} + c_{i_\downarrow}^\dagger \, c_{a_\downarrow}.
\end{align}
\end{subequations}
where $G(L_2)$ is a polynomial we will specify shortly.  We will make the energy stationary with respect to the four distinct single-excitation amplitudes:
\begin{equation}
0 = \frac{\partial E}{\partial t_i^a}
  = \frac{\partial E}{\partial u_i^a}
  = \frac{\partial E}{\partial z^i_a}
  = \frac{\partial E}{\partial v^i_a}.
\end{equation}
Note the strong resemblance to coupled-cluster doubles, which has
\begin{subequations}
\begin{align}
T &= T_2 = \frac{1}{2} \, \sum t_{ij}^{ab} \, E_a^i \, E_b^j,
\\
Z &= Z_2 = \frac{1}{2} \, \sum z^{ij}_{ab} \, E^a_i \, E^b_j,
\end{align}
\end{subequations}
though here we have a more complicated left-hand state and, crucially, $K_2$ and $L_2$ have factorizable amplitudes while $T_2$ and $Z_2$ in general do not.

There are two general stategies we might pursue for the bra polynomial $G(L_2)$.  One is to make $G(x) = F(x)$ in analogy with extended coupled cluster theory.\cite{Arponen1983,Piecuch2006}  The second is to try to match this projective PHF energy expression to the variational one, in which case we want to have schematically
\begin{equation}
G(L_2) \, F^{-1}(K_2) \approx \frac{F(K_2^\dagger)}{\langle \mathrm{SUHF} | \mathrm{SUHF}\rangle}.
\end{equation}
While we cannot in general enforce this condition exactly, we \textit{can} enforce it on average.  In other words, we can adjust the coefficients $c_n$ in $G(L_2) = \sum c_n \, L_2^n$  by imposing conditions like 
\begin{equation}
\langle \mathrm{RHF} | G(L_2) \, F^{-1}(K_2) |n\rangle = \frac{\langle \mathrm{RHF} | F(K_2^\dagger) | n\rangle}{\langle \mathrm{SUHF}|\mathrm{SUHF}\rangle},
\end{equation}
where $|n\rangle$ stands for $n$-tuply excited determinants.  This leads to a set of linear equations for the coefficients $c_n$.  We consider both approaches.

\begin{figure}[t]
\includegraphics[width=\columnwidth]{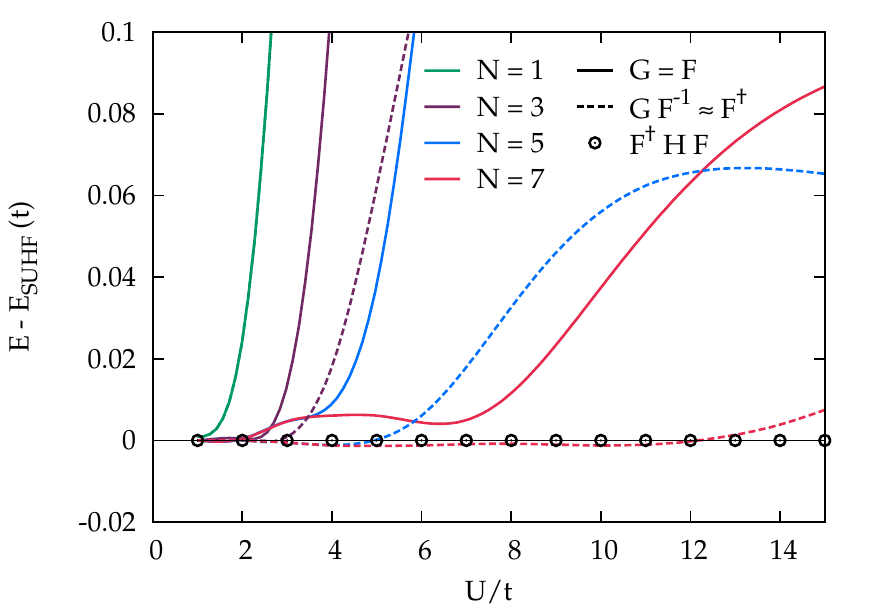}
\caption{Energies relative to SUHF in the half-filled 14-site Hubbard Hamiltonian.  Solid lines indicate $G(x) = F(x)$ and dashed lines indicate we have fit $G(L_2) \, F^{-1}(K_2) \sim F(K_2^\dagger)$.  The label ``N'' indicates that $G(x)$ has been truncated at $L_2^N$.  Dots denote the variational energy expression.
\label{Fig:Hubbard14}}
\end{figure}

We should say a few words about computational complexity.  For connected (non-factorizable) double-excitation operators, evaluating either expectation values of $F(K_2^\dagger) \, H \, F(K_2)$ or $G(L_2) \, \bar{H}_\mathrm{PHF}$ would be prohibitively expensive unless we truncated the polynomials $F(x)$ and $G(x)$ to low order, but a key difference between our approach and CC theory is that our energy expression does not truncate ; only similarity transformations generated by exponentials lead to terminating series (at $\mathcal{O}(T^4)$ for a two-body Hamiltonian).  However, as we shall discuss in detail in a subsequent manuscript, the evaluation of terms even with high powers of $K_2$ or $L_2$ is feasible, essentially because $K_2$ and $L_2$ factorize, and yields relatively simple expressions with polynomial computational cost.  Nevertheless, we shall here show results where we manually truncate $G(L_2)$, so one can assess convergence toward PHF.

Figure \ref{Fig:Hubbard14} shows energies relative to SUHF for the 14-site Hubbard Hamiltonian.  The dot circles represent energies obtained via Eq. (\ref{Eqn:FDagHF}), numerically demonstrating the exactness of the sinh polynomial of Eq. (\ref{Eqn:Sinh}).  It is clear that a low-order truncation of $G(L_2)$ is inadequate; although not shown, the same is true for $F(K_2^\dagger)$.  Second, with $G = F$, our projective approach replicates standard PHF to a reasonable extent.  Finally, when we optimize the polynomial $G$ to match the variational expectation value, we reproduce PHF even better.  The remaining small discrepancies between our projective approach and the exact SUHF are presumably because we cannot make $G(L_2) \, F^{-1}(K_2)$ match $F(K_2^\dagger)$ exactly.  Note that in the Hubbard Hamiltonian, $T_1 = 0$ by symmetry, so these results are obtained purely with $K_2$ (or, if one prefers, $U_1$).

\begin{figure}[t]
\includegraphics[width=\columnwidth]{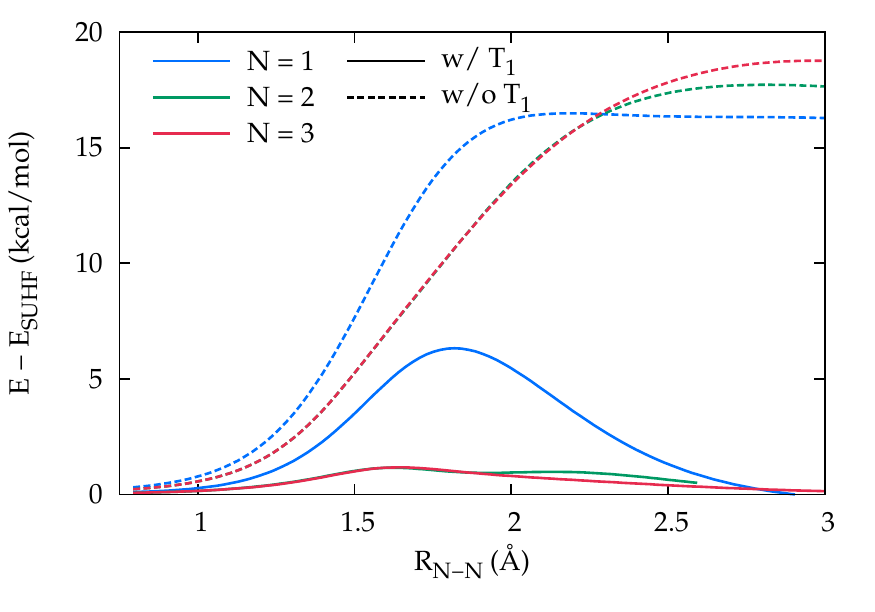}
\caption{Energies relative to SUHF in the dissociation of N$_2$, using the cc-pVDZ basis.  We fit $G \, F^{-1} \sim F^\dagger$ with $G$ truncated to order $N$; solid and dashed lines show results with and without the inclusion of $T_1$.
\label{Fig:N2DZ}}
\end{figure}

We can see the importance of single excitations in Fig. \ref{Fig:N2DZ} where we show results for the dissociation of N$_2$ in the cc-pVDZ basis with an optimized polynomial $G(L_2)$.  As in the Hubbard Hamiltonian we converge toward SUHF as we increase the degree of the polynomial $G$, but only when we include $T_1$.  This is of course what one would expect on the basis of Eq. (\ref{Eqn:Thouless}).

We wish to emphasize that while one can truncate $F(K_2^\dagger)$ or $G(L_2)$, this truncation only really converges at order $N$, where $N$ is the number of strongly correlated pairs.  Thus, in the half-filled 14-site Hubbard Hamiltonian where there are 14 strongly correlated electrons, we must in general retain terms all the way up to $L_2^7$, where in $N_2$ where there are six strongly correlated electrons we need keep terms only up to $L_2^3$.  This is in analogy with the conventional wisdom in standard coupled cluster theory.  Unlike in standard coupled cluster, retaining terms of high order is straightforward because $K_2$ and $L_2$ are ultimately obtained from \textit{single} excitation amplitudes.

Finally, let us reiterate one last time what we wish to demonstrate in this work.  Previous work on polynomial similarity transformations\cite{Scuseria2016} showed how one could interpolate between the number projected BCS and coupled cluster forms of wave function by writing them both in the same language.  Rather than interpolating, however, we would like to combine coupled cluster theory and symmetry-projected mean-field methods in a more sophisticated and presumably more correct wave function, a task made more difficult by the very different natures of the two theories (but see Ref. \onlinecite{Duguet} for a possible solution in the broken symmetry basis).  In this manuscript, we have shown how to use a coupled-cluster-like formalism to solve for the energy and wave function of projected Hartree-Fock theory.  By so doing, we set the stage for the explicit wave function-based combination of PHF and CC, thereby obtaining the best of both worlds.

\textbf{Acknowledgments:} 
This work was supported by the U.S. Department of Energy, Office of Basic Energy Sciences, Computational and Theoretical Chemistry Program under Award No.DE-FG02-09ER16053. G.E.S. is a Welch Foundation Chair (C-0036).  We thank Jorge Dukelsky and Matthias Degroote for helpful discussions.  Jinmo Zhao's \texttt{pyslata} code was helpful for debugging our energy expression.

\bibliography{PoST}

\end{document}